**Comprehensive OOS Evaluation of Predictive Algorithms with Statistical Decision Theory**

Jeff Dominitz
Department of Economics, Justice, and Society – NORC at the University of Chicago

and

Charles F. Manski
Department of Economics and Institute for Policy Research, Northwestern University

Abstract

We argue that comprehensive out-of-sample (OOS) evaluation using statistical decision theory (SDT) should replace the current practice of K-fold and Common Task Framework validation in machine learning (ML) research on prediction. SDT provides a formal frequentist framework for performing comprehensive OOS evaluation across all possible (1) training samples, (2) populations that may generate training data, and (3) populations of prediction interest. Regarding feature (3), we emphasize that SDT requires the practitioner to directly confront the possibility that the future may not look like the past and to account for a possible need to extrapolate from one population to another when building a predictive algorithm. For specificity, we consider treatment choice using conditional predictions with alternative restrictions on the state space of possible populations that may generate training data. We discuss application of SDT to the problem of predicting patient illness to inform clinical decision making. SDT is simple in abstraction, but it is often computationally demanding to implement. We call on ML researchers, econometricians, and statisticians to expand the domain within which implementation of SDT is tractable.

We have benefitted from the opportunity to present this work in seminars at Northwestern University and the University of Chicago, and at conferences at Brown University and Cornell University. We have further benefitted from the comments of reviewers on an earlier draft.

## 1. Introduction

As hopes and fears related to artificial intelligence (AI) continue to grow, it is important to recognize that the impact of any AI system will always be tied to the quality of the predictive algorithms on which the AI is based. Throughout the 20th century, evaluation of predictions made with sample data was primarily the domain of statisticians and econometricians, who use frequentist or Bayesian statistical theory to propose and evaluate probabilistic prediction methods. In the 21st century, prediction is increasingly performed by researchers in computer science, who view prediction methods as computational algorithms and who rarely use statistical theory to assess the algorithms. Instead, these researchers commonly perform *out-of-sample* (OOS) tests of predictive accuracy.

In an influential early article, Breiman (2001a) argued for this approach, writing (p. 201): "Predictive accuracy on test sets is the criterion for how good the model is." He asserted that OOS evaluation is close to theory-free, stating (p. 205): "the one assumption made in the theory is that the data is drawn i.i.d. from an unknown multivariate distribution." Although Breiman did not state it explicitly, we conjecture that he had in mind settings where the data are drawn from the distribution of prediction interest, not from just *any* unknown distribution.

Over the past two decades, OOS evaluation has often been applied even without this basic assumption. Using the now familiar term *machine learning* (ML) to describe prevalent prediction practices in computer science, Taddy (2019) remarked on the wide adoption of the approach advocated by Breiman, writing (p.70): "Machine learning's wholesale adoption of OOS validation as the arbitrator of model quality has freed the ML engineer from the need to *theorize* about model quality."

The term *machine learning* has been used in multiple ways, with occasional lack of clarity. Considered broadly, ML means any computational procedure using data to make predictions. This broad usage encompasses the host of parametric and nonparametric probabilistic prediction methods developed by statisticians and econometricians over the past 150 years, studied through the lens of frequentist or Bayesian statistical theory. The recent usage in computer science is narrower, focusing on certain nonparametric methods (e.g., random forests and deep neural networks), whose predictive properties have been assessed by OOS evaluation. In this paper, we mainly use the term in the narrow sense, as does Taddy.[1]

Taddy described OOS evaluation this way (p. 70):

> "Out-of-sample validation is a basic idea: you choose the best model specification by comparing predictions from models estimated on data that was not used during the model 'training' (fitting). This can be formalized as a cross-validation routine: you split the data into K 'folds,' and then K times fit the model on all data but the K[th] fold and evaluate its predictive performance (e.g., mean squared error or misclassification rate) on the left-out fold. The model with optimal average OOS performance (e.g., minimum error rate) is then deployed in practice."

Cross-validation (CV) has long been used by statisticians to choose tuning parameters for prediction with nonparametric and semiparametric regression methods. Statisticians have used frequentist asymptotic statistical theory to evaluate prediction methods with cross-validated tuning parameters—studying their consistency, rates of convergence, and limiting distributions (e.g., Zhang, 1993). The use of CV in OOS evaluation described by Tandy differs. Here one evaluates predictive performance on alternative subsets of the realized training sample, without considering how the method would perform across repeated draws of the training sample. Thus, K-fold OOS CV does not use frequentist statistical theory. Note, however, that Bates, Hastie, and Tibshirani (2024) have recently offered a frequentist statistical perspective on OOS CV.

Rather than use a K-fold split of the training data to form multiple validation samples, OOS evaluation is sometimes performed in the so-called 'Common Task Framework' (CTF). Here, predictive accuracy is measured in some pre-specified testing data that may differ systematically from the training sample. Donoho (2017) described the CTF as follows:

[1] An apt example of lack of clarity in usage concerns *empirical risk minimization (ERM)* as studied by Vapnik and others in the latter part of the 1990s; see Vapnik (1999, 2000). Computer scientists frequently embrace this methodology within the domain of machine learning. However, it firmly is a development of frequentist statistical theory, grounded in limit theorems regarding consistency and rates of convergence. Indeed, ERM is a direct descendant of classical statistical analysis of the method of moments, M-estimation, and minimum distance estimation. All of these methods apply what Goldberger (1968) called the *analogy principle*, using the empirical distribution of a training sample to estimate a population distribution under the assumption of random sampling, relying on a Law of Large Numbers for their basic convergence properties. Computer scientists do not ordinarily refer to these classical statistical methods as machine learning, yet they apply this term to empirical risk minimization.

> "An instance of the CTF has these ingredients: (a) A publicly available training dataset involving, for each observation, a list of (possibly many) feature measurements, and a class label for that observation. (b) A set of enrolled competitors whose common task is to infer a class prediction rule from the training data. (c) A scoring referee, to which competitors can submit their prediction rule. The referee runs the prediction rule against a testing dataset, which is sequestered behind a Chinese wall. The referee objectively and automatically reports the score (prediction accuracy) achieved by the submitted rule. All the competitors share the *common task* of training a prediction rule which will receive a good score; hence the phase *common task framework*. A famous recent example is the Netflix Challenge, where the common task was to predict Netflix user movie selections."

CTF evaluation is the same as 1-fold CV when the testing dataset behind the wall is a hold-out subsample of the training sample. However, the designated common task often is to predict an outcome in a testing dataset not generated from the training sample. The practice of holding competitions with awards made using the CTF is remote from the concern of our paper, which is decision making by a single agent.

The intuitive appeal of K-fold and CTF OOS evaluation has aided computer scientists in their efforts to market ML predictive algorithms to medical clinicians, judges and police, firm managers, and others who may lack expertise in statistical theory but who feel that they can appraise prediction accuracy heuristically. However, it should be obvious that these types of OOS evaluation cannot yield generalizable lessons. We acknowledge that innovation without theory has sometimes been productive throughout human history, but we are concerned that it sometimes leads society astray.

Caution was expressed early on by Cox (2001), whose published Comment on Breiman (2001a) distinguished between cases where (p. 216-17): "prediction is localized to situations directly similar to those applying to the *[training]* data" and those where "prediction is under quite different conditions from the data." On the former, Cox concluded that the algorithmic "empirical black-box approach" may be preferable to explicit modelling of the data generating process, whereas, in the latter case "prediction, always hazardous, without some understanding of underlying process and linking with other sources of information, becomes more and more tentative." In his Rejoinder, Breiman (p. 227) replied to the latter statement by stating, simply, "I agree."

More recently, Taddy (2019) wrote (p. 62): "Machine learning can do fantastic things, but it is basically limited to predicting a future that looks mostly like the past." Poggio and Fraser (2024) caution that the popular ML approach of fitting training data to deep neural networks should be expected to yield good predictions of an outcome y conditional on covariates x only if the true best predictor of interest (often the conditional mean function) is "compositionally sparse;" that is, if it has a structure similar to that of deep neural network functions. We share the important concerns of Cox, Taddy, and Poggio and Fraser.

Efron (2020), in an article contrasting the perspectives of statisticians and computer scientists, wrote (p. S49): "In place of theoretical criteria, various prediction competitions have been used to grade algorithms in the so-called 'Common Task Framework.' . . . None of this is a good substitute for a so-far nonexistent theory of optimal prediction." We agree with Efron that the prediction competitions of the CTF are not a satisfactory way to evaluate prediction methods. However, we sharply disagree with Efron's second statement. To reiterate what Manski (2023) observed recently (p. 648):

> "[Efron] was not correct when he stated that a theory of optimal prediction is 'so-far nonexistent'. Wald (1939, 1945, 1950) considered the general problem of using sample data to make decisions. He posed the task as choice of a *statistical decision function*, which maps potentially available data into a choice among the feasible actions. His development of statistical decision theory provides a broad framework for decision making with sample data, yielding optimal decisions when these are well-defined and proposing criteria for reasonable decision making more generally."

Wald recommended ex ante (frequentist) evaluation of statistical decision functions as *procedures* to be applied as a sampling process is engaged repeatedly to draw independent data samples. Statistical decision theory (SDT) measures the mean performance of a prediction method across all possible training samples, when the objective is to predict outcomes in a population that may possibly differ from the one generating the data in the training sample.

We emphasize that SDT performs *ex ante* evaluation. This is directly at odds with the standard practice of ML researchers to perform *ex post* evaluation, after observation of a particular training sample and without regard to other training samples that might have been drawn. Ex ante evaluation is the standard practice in frequentist statistical theory, which evaluates estimates by their sampling distributions, using criteria including consistency, unbiasedness, and efficiency.

SDT is remote from the types of OOS evaluation currently practiced by ML researchers, but it is not remote conceptually. Indeed, SDT provides a formal framework for performing what we shall term *comprehensive OOS* evaluation. SDT performs OOS evaluation across all possible (1) training samples, (2) populations that may generate training data, and (3) populations of prediction interest. We think that feature (3) is particularly important. SDT requires the practitioner to directly confront the possibility that the future may not look like the past and to account for a possible need to extrapolate from one population to another when building a predictive algorithm.

Even in the best-case scenario where the future will credibly look like the past—i.e., no extrapolation problem—SDT provides a framework for OOS evaluation that is scientifically more rigorous than the current standard practice of K-fold and CTF validation. When the training sample is known to be drawn i.i.d. from the distribution of prediction interest, one must recognize that this sample is just one draw among all possible samples from this population distribution. With comprehensive OOS evaluation, the performance of the predictive algorithm is assessed across all samples on which it could be trained, rather than just the one sample on which it actually is trained. Thus, SDT embraces the paradigm of frequentist statistical theory.

ML researchers often motivate their versions of OOS evaluation by stating that it protects against drawing misleading conclusions from prediction performance on the training sample, which may be unrealistically high due to so-called "overfitting" the data. Concern with overfitting does not arise in the Wald framework because it evaluates performance across all feasible samples, not a particular training sample. Whereas some may believe that the so-called "big data" revolution renders attention to sampling variation unnecessary, the desire to build ML models with high-dimensional covariates gives rise to the same finite sample concerns that SDT was developed to address. Even if the total sample size is orders of magnitude larger than Wald may have envisioned in the 1940s, statistical imprecision is an important consideration when conditioning on covariates whose distribution has sufficiently large support.

In this paper, we argue that comprehensive OOS evaluation performed using SDT should replace the current practice of K-fold and CTF validation in ML research. The argument is simple to make in abstraction. As we explain in Section 2, the Wald framework is general, transparent, and intellectually attractive. In principle, it enables comparison of essentially all predictive algorithms, the only caveat being satisfaction of weak mathematical regularity conditions. It enables comparison of alternative sampling processes generating training data. It uses no asymptotic approximations when interpreting the training data. Further, the population of interest in prediction may differ from that generating the training data.

The argument is more challenging to make in practice. Application of SDT is easy in some important settings, but it is often computationally demanding. Computation commonly requires numerical methods to find approximate solutions. Moreover, SDT requires the practitioner to specify a decision criterion. Among the criteria that have been studied, we find minimax regret particularly appealing. A statistical decision function (SDF) with small maximum regret is uniformly near-optimal across all possible populations generating training data and populations of prediction interest. In the terminology of SDT, a possible pair of such populations is a *state of nature.* The set of all possible populations is the *state space*. Maximum regret is computed across the state space.

To go beyond abstraction, Sections 3 and 4 consider treatment choice using a prediction of an outcome y conditional on a specified covariate value x. To focus on the problem of statistical imprecision, we suppose that the population of prediction interest is the same as that generating the training data. The central issue is specification of cross-covariate restrictions on conditional outcome distributions, which enable outcome data associated with covariate values $x' \neq x$ to be informative about $P(y|x)$.

Statistical and econometric theory has long studied settings in which cross-covariate restrictions are imposed by assuming all conditional distributions, or their means, lie in a specified finite-dimensional set (parametric regression) or a specified space of suitable smooth functions (nonparametric regression). The ML literature using deep neural networks as predictor functions has recently favored assumptions that conditional probability distributions or means are compositionally sparse, which Poggio and Fraser (2024) described as (p. 1): "a key principle underlying successful learning architectures." Yet other cross-covariate restrictions on conditional distributions are specified in the bounded-variation assumptions studied by Manski (2023). We do not take a stand favoring any one class of restrictions on the state space over another—the choice should be application specific. Our theme is that SDT may in principle be used to perform comprehensive OOS evaluation, however a researcher chooses to restrict the state space.

We discuss the important concrete problem of predicting patient illness to inform clinical decision making. We summarize the history of probabilistic prediction by biostatisticians and computer scientists. We then focus on the problem of choice between surveillance and aggressive treatment of a patient whose illness status is not observed but can be predicted probabilistically.

Past Econometric Research Applying Statistical Decision Theory

Readers wanting to learn about the early development of statistical decision theory may want to begin with the highly informative monographs of Ferguson (1967) and Berger (1985). Econometric research applying SDT began in the mid-20th century (e.g., Sawa and Hiromatsu, 1973), but it has become common more recently. Much of the focus has been on minimax-regret treatment choice with data from classical randomized trials or quasi-experiments, where the distribution of treatment response is point-identified and the sole concern is statistical imprecision. Contributions include Manski (2004, 2005, 2019), Hirano and Porter (2009, 2020), Stoye (2009), Manski and Tetenov (2016, 2019, 2021), Kitagawa and Tetenov (2018), and Mbakop and Tabord-Meehan (2021).[2]

Attention is increasingly being given to treatment choice with data from trials with imperfect internal validity or from observational studies, where identification is the dominant concern with large sample size and joint attention to identification and statistical imprecision is essential when sample size is small. Contributions include Athey and Wager (2021), Manski (2000, 2007, 2021, 2023), Montiel Olea, Qiu, and Stoye (2024), Stoye (2012), and Yata (2023). Dominitz and Manski (2017, 2022) studied prediction under square loss with nonrandomly missing data.

---

[2] With the exception of Hirano and Porter (2009, 2020), these studies apply Wald's finite-sample theory. Hirano and Porter use the local asymptotic theory of Hajek and LeCam to circumvent the computational complexity often encountered when implementing the Wald finite-sample theory. At present, little is known about the adequacy of the the Hajek-LeCam theory to approximate the finite-sample settings that occur in applied research.

## 2. The Statistical Decision Theory Perspective on Prediction

When Wald formally considered the general problem of using sample data to make decisions, this included the use of data to make predictions. We present the general structure of SDT in Section 2.1. Section 2.2 explains how SDT views prediction—in particular, SDFs that use predictions to make decisions. Section 2.3 juxtaposes statistical decision theory and research on robust decisions. This exposition draws on Manski (2021, 2023, 2024).

### 2.1. Statistical Decision Theory in Abstraction

Wald utilized the standard decision-theoretic framing of the choice problem of a decision maker (DM) who must choose an action yielding loss that depends on an unknown state of nature. The DM specifies a state space listing the states considered possible. The DM wants to minimize loss, but must choose without knowing the true state.

To begin, Section 2.1.1 presents this problem in a setting where the DM does not observe sample data. Section 2.1.2 explains how Wald generalized this setting by supposing that the DM observes sample data that may be informative about the true state. Section 2.1.3 discusses computation.

Wald's first major contribution was to recast the hypothesis-testing problem previously posed by Neyman and Pearson (1928, 1933) as a decision problem in which one must choose between two feasible actions. ML researchers now refer to this type of decision as a binary *classification problem*. We summarize Wald's formalization of the classification problem in Appendix A,

2.1.1. Decisions Without Sample Data

Consider a DM who faces a predetermined choice set C and believes that the true state $s^*$ lies in state space S. The state space may be finite-dimensional (parametric) or larger (nonparametric). Objective function $L(\cdot, \cdot)$: $C \times S \to R^1$ maps actions and states into loss. The DM ideally would minimize $L(\cdot, s^*)$ over C, but the DM does not know $s^*$.

It is generally accepted that choice should respect dominance. Action $c \in C$ is weakly dominated if there exists a $d \in C$ such that $L(d, s) \leq L(c, s)$ for all $s \in S$ and $L(d, s) < L(c, s)$ for some $s \in S$. To choose among undominated actions, decision theorists have proposed various ways of using $L(\cdot, \cdot)$ to form functions of actions alone, which can be optimized.[3]

Wald (1945) studied choice when the DM places a subjective probability distribution $\pi$ on the state space, averages state-dependent loss with respect to $\pi$, and minimizes subjective average loss $\int L(c, s)d\pi$ over C. The criterion solves

$$(1) \qquad \min_{c \in C} \int L(c, s)d\pi.$$

Wald (1945, 1950) considered choice when the DM does not place a subjective distribution on the state space. In this setting, he studied minimax choice, which selects an action that works uniformly well over all of S in the sense of minimizing the maximum loss attainable across S. The minimax criterion is

$$(2) \qquad \min_{c \in C} \max_{s \in S} L(c, s).$$

[3] In principle, one should only consider undominated actions, but it often is difficult to determine which actions are undominated. Hence, in practice it is common to optimize over the full set of feasible actions. We define decision criteria accordingly. We use max and min notation, without concern for the subtleties that sometimes make it necessary to use sup and inf operations.

Savage (1951), in a book review of Wald (1950), suggested a different formalization of the idea of selecting an action that works uniformly well over all of S. This formalization, which has become known as the minimax-regret (MMR) criterion, solves the problem

$$(3) \quad \min_{c \in C} \max_{s \in S} [L(c, s) - \min_{d \in C} L(d, s)].$$

Here $L(c, s) - \min_{d \in C} L(d, s)$ is the *regret* of action c in state s—i.e., the increment to loss in state s arising from making choice c rather than the optimal choice in that state. The true state being unknown, one evaluates c by its maximum regret over all states and selects an action that minimizes maximum regret. The maximum regret of an action measures its maximum distance from optimality across states.[4]

2.1.2. Statistical Decision Problems

Statistical decision problems add to the above structure by supposing that the DM observes data generated by some sampling distribution. Knowledge of the sampling distribution is generally incomplete. To express this, one extends state space S to list the feasible sampling distributions, denoted $(Q_s, s \in S)$. Let $\Psi_s$ denote the sample space in state s; $\Psi_s$ is the set of samples that may be drawn under distribution $Q_s$. The literature typically assumes that the sample space does not vary with s and is known. We assume this and denote the sample space as $\Psi$. Then a statistical decision function, $c(\cdot)$: $\Psi \rightarrow C$, maps the sample data into a chosen action. Henceforth, $\psi \in \Psi$ is a possible realization of the sample data; that is, a possible training sample.

Wald's concept of a statistical decision function embraces all mappings [data → action]. SDF $c(\cdot)$ is a deterministic function after realization of the sample data, but it is a random function ex ante. Hence, the loss achieved by $c(\cdot)$ is a random variable ex ante. Wald's theory evaluates the performance of $c(\cdot)$ in state s by $Q_s\{L[c(\psi), s]\}$, the ex-ante distribution of loss that it yields across realizations $\psi$ of the sampling process.

[4] Criteria (1)—(3) have become the most prominent studied in decision theory, but they are not alone. Hurwicz (1951) suggested minimization of a weighted average of worst and best possible outcomes. Rather than assert a complete subjective distribution on the state space or none, a DM might assert a partial subjective distribution, placing lower and upper probabilities on states. One then might minimize maximum subjective average loss or minimize maximum subjective average regret. These criteria combine elements of averaging across states and concern with uniform performance across states. It appears that this idea was first suggested by Hurwicz (1951). The idea was later taken up by statistical decision theorists. See Berger (1985) and Walley (1990).

It remains to ask how DMs might compare the loss distributions yielded by different SDFs. DMs want to minimize loss, so it seems self-evident that they should prefer SDF $d(\cdot)$ to $c(\cdot)$ in state s if $Q_s\{L[d(\psi), s]\}$ is stochastically dominated by $Q_s\{L[c(\psi), s]\}$. It is less obvious how they should compare SDFs whose loss distributions do not stochastically dominate one another.

Wald proposed measurement of the performance of $c(\cdot)$ in state s by its expected loss across samples; that is, $E_s\{L[c(\psi), s]\} \equiv \int L[c(\psi), s]dQ_s$. He used the term *risk* to denote the mean performance of an SDF across samples. An alternative that has drawn only slight attention measures performance by quantile loss (Manski and Tetenov, 2023).

Not knowing the true state, a DM evaluates $c(\cdot)$ by the expected loss vector $(E_s\{L[c(\psi), s]\}, s \in S)$. Using the term *inadmissible* to denote weak dominance when evaluating performance by risk, Wald recommended elimination of inadmissible SDFs from consideration. As in decisions without sample data, there is no clearly best way to choose among admissible SDFs. SDT has mainly studied the same criteria as has decision theory without sample data. Let $\Gamma$ be a specified set of SDFs, each mapping $\Psi \rightarrow C$. The statistical versions of criteria (1), (2), and (3) are

$$(4) \qquad \min_{c(\cdot) \in \Gamma} \int E_s\{L[c(\psi), s]\}\, d\pi,$$

$$(5) \qquad \min_{c(\cdot) \in \Gamma} \max_{s \in S} E_s\{L[c(\psi), s]\},$$

$$(6) \qquad \min_{c(\cdot) \in \Gamma} \max_{s \in S} \left( E_s\{L[c(\psi), s]\} - \min_{d \in C} L(d, s)\right).$$

Each of criteria (4) – (6) has been deemed reasonable by some decision theorists, but each has also drawn criticism. Minimization of subjective average loss (4) (aka Bayes risk) may be appealing if one has a credible basis to place a subjective probability distribution on the state space, but not otherwise. Concern with specification of priors motivated Wald to study the minimax criterion (5). However, minimax was criticized by Savage (1951) as 'ultrapessimistic.' Manski (2004, 2021) argued for the MMR criterion (6), observing that an SDF with small maximum regret is uniformly near-optimal across all states.

Whichever of criteria (4)—(6) one uses, SDT performs a comprehensive OOS evaluation of an SDF. In each state of nature s, the expected loss $E_s\{L[c(\psi), s]\}$ of SDF $c(\cdot)$ measures its performance across all possible training samples. The state-dependent vector $\{E_s\{L[c(\psi), s]\}, s \in S\}$ of expected loss measures performance across all possible populations that may have generated the training sample and all possible populations of decision interest. Direct measurement of performance across all possible populations replaces any need for test samples to be drawn.

Recall that, when arguing for his narrow form of OOS evaluation, Breiman (2001a) stated that "the one assumption made in the theory is that the data is drawn i.i.d. from an unknown multivariate distribution." This assumption is unnecessary in SDT. Any sampling distribution can generate the data.

2.1.3. Computation

Subject to regularity conditions ensuring that the relevant expectations and extrema exist, problems (4) – (6) offer criteria for decision making with sample data that are broadly applicable in principle. The primary challenge is computational. Problems (4) − (6) have tractable analytical solutions only in certain cases. Computation commonly requires numerical methods to find approximate solutions.

Expected loss $E_s\{L[c(\psi), s]\}$ typically does not have an explicit form, but it can be well-approximated by Monte Carlo integration. One draws repeated values of $\psi$ from distribution $Q_s$, computes the average value of $L[c(\psi), s]$ across the values drawn, and uses this to estimate $E_s\{L[c(\psi), s]\}$. Monte Carlo integration can also be used in criterion (4) to approximate the subjective average of expected loss.

The main computational challenges are determination of the extrema across actions in problem (6), across states in problems (5) − (6), and across SDFs in problems (4) − (6). Solution of min $_{d \in C}$ L(d, s) in (6) is often straightforward but sometimes difficult. Finding extrema over S must cope with the fact that the state space commonly is uncountable. In applications where the quantity to be optimized varies smoothly over S, a simple approach is to compute the extremum over a suitable finite grid of states.

The most difficult computational challenge usually is to optimize over the feasible SDFs. No generally applicable approach is available. Hence, applications of SDT necessarily proceed case-by-case. It may not be tractable to find the best feasible SDF, but one often can evaluate the performance of relatively simple SDFs that researchers use in practice.

## 2.2. Prediction-Based SDFs

We have observed that Wald's concept of an SDF embraces all mappings [data → action]. This very broad domain includes SDFs that make predictions and use the predictions to make decisions. These have the form [data → prediction → action], first making a prediction and then using the prediction to make a decision. There seems to be no accepted term for such SDFs, so we call them *prediction-based* SDFs.

To formalize prediction-based SDFs, we first need to formalize the concept of prediction and embed it in a decision problem. To do this, we bring to bear the venerable idea of probabilistic conditional prediction. This idea postulates a population characterized by a joint probability distribution P(y, x), where (y, x) takes values in some set Y × X. A member is drawn at random from the sub-population with a specified value of x. Then the conditional distribution P(y|x) provides the ideal probabilistic prediction of y conditional on x.

We embed prediction in a decision problem by considering settings in which a value of x is specified and the state space contains a set of feasible conditional distributions P(y|x). Then $P^*(y|x)$ is the unknown distribution of prediction interest. The DM wants to minimize L[·, $P^*(y|x)$] over C, but does not know $P^*(y|x)$. One faces a statistical decision problem if one observes data ψ drawn from some sampling process.

In a setting of this type, an SDF is prediction-based if the DM uses the data $\psi$ to form an estimate of $P^*(y|x)$ and acts as if the estimate is accurate. Let $f(\cdot)$: $\Psi \to S$ be the predictor function used to estimate $P^*(y|x)$. The chosen action with this SDF is $d(\psi) = \text{argmin}_{\, c \in C}\, L[c, f(\psi)]$.

Rather than estimate the entire conditional distribution $P^*(y|x)$, researchers often use sample data to form a point prediction of outcome y. Let $p(\cdot)$: $\Psi \to Y$ denote a predictor function that generates a point prediction. The chosen action with an SDF of this type acts as if $P^*(y|x)$ is degenerate, with all its mass at the outcome value $p(\psi)$. Thus, the SDF is $d(\psi) = \text{argmin}_{\, c \in C}\, L[c, p(\psi)]$.

Important special cases occur when the choice set C coincides with the set Y of potential outcomes. Let y be real-valued and let loss be MSE when using c to predict y. Then C = Y = R and

$$(7) \qquad L[c, P(y|x)] = E[(y - c)^2|x] = \int (y - c)^2 dP(y|x).$$

Or y may be binary, and loss may be the misclassification rate (MCR) when using c to predict y. Then C = Y = {0, 1} and

$$(8) \qquad L[c, P(y|x)] = P(y \neq c|x) = \int 1[y \neq c] dP(y|x).$$

Observe that $1[y \neq c] = (y - c)^2$ when C = {0, 1}, but not otherwise. See Appendix A for further discussion.

From the perspective of SDT, the performance of a prediction-based SDF should be evaluated in the same manner as any SDF. Thus, one should first determine if the SDF is undominated. If so, one may evaluate it by the subjective expected loss it yields, the maximum loss it yields, by its maximum regret, or by some other reasonable criterion.

2.3. Robust Decisions

To conclude this introduction to statistical decision theory, we explain its relationship to research on *robust decisions*. The field of robust statistics began with study of prediction when data contamination makes the probability distribution generating the training data lie in a neighborhood of the distribution of prediction interest (e.g., Huber, 1964, 1981; Hampel et. al., 1986; Horowitz and Manski, 1995). Subsequently, engineers, economists, and computer scientists have studied mathematically similar but distinct problems of robust decision making in which a researcher specifies a probabilistic model space and assumes that the true probability distribution generating outcomes lies in a neighborhood of the model space (e.g., Zhou, Doyle, and Glover, 1996; Hansen and Sargent, 2008). Blanchet et al. (2024) refers to this as *distributional shift*.

Study of robust decisions has proceeded in a different manner than statistical decision theory. Rather than initially specify a state space that lists all possible states of nature, the researcher initially poses a model space. The model specifies a single state or a relatively small set of states, often a finite-dimensional family. Having posed the model space, a researcher may be concerned that it does not contain the true state; that is, the model may not be correct. To recognize this possibility, the researcher enlarges the model space locally, using some metric to generate a neighborhood thereof. He then acts as if the locally enlarged model space is correct. Watson and Holmes write (p. 465): "We then consider formal methods for decision making under model misspecification by quantifying stability of optimal actions to perturbations to the model within a neighbourhood of [the] model space."

Although research on robust decisions differs procedurally from statistical decision theory, one can subsume the former within the latter if one considers the locally enlarged model space to be the state space. It is unclear how often this perspective characterizes what researchers have in mind. Articles often do not state explicitly whether the constructed neighborhood of the model space encompasses all states that authors deem sufficiently feasible to warrant consideration. The models specified in robust decision analyses often make strong assumptions and generated neighborhoods often relax these assumptions only modestly.

## 3. Conditional Prediction with Cross-Covariate Restrictions on the State Space

In principle, SDT may be used to perform comprehensive OOS evaluation of any of the huge variety of probabilistic conditional prediction methods that have been developed from the late 1800s onward by statisticians, econometricians, and ML researchers. In practice, SDT has rarely been used to evaluate conditional prediction methods by researchers in these fields and applied practitioners. Prediction has mainly been studied under the assumption that the training data are a random sample drawn from the distribution P(y, x) of prediction interest, or at least that P(y|x) generates the training realizations of y conditional on x. The literature on Bayesian prediction predominately uses the *conditional Bayes* paradigm. This centers attention on the posterior predictive distribution, computed after training data are observed, not on ex ante Bayes risk as in SDT. There has been little minimax or MMR evaluation of conditional prediction methods; isolated cases of MMR evaluation of linear regression are Sawa and Hiromatsu (1973) and Chamberlain (2000). Research on nonparametric regression has mainly studied asymptotic properties of estimates when the covariate distribution has positive density in a neighborhood of a value of interest and the conditional expectation E(y|x) varies smoothly with x in a local sense, such as differentiability. Donoho et al. (1995) reviews many findings. An isolated instance of MMR evaluation of kernel nonparametric regression is Manski (2023), which bounds the global rather than local variation of E(y|x) with x. See Section 3.2 below for discussion.

Whereas statisticians and econometricians have used statistical theory to study prediction methods, ML researchers have largely performed heuristic K-fold and CTF OOS validation. Despite the absence of theory, ML researchers assert that K-fold and CTF validation exercises show that certain favored methods commonly "work" in practice. Yet the reasons why these methods "work" and the circumstances in which they "work" continue to be controversial. This is clear from a widely cited report of Bommasani et al. (2021), who concluded (p. 1): "Despite the impending widespread deployment of foundation models, we currently lack a clear understanding of how they work, when they fail, and what they are even capable of."

We are aware of only a few frequentist studies of ML methods. Schmidt-Hieber (2020) studied the rates of convergence of deep neural network predictors when conditional means are compositionally sparse. Bartlett et al. (2021) used asymptotic statistical theory to study the predictive accuracy of deep learning methods that involved "benign overfitting." Bates, Hastie, and Tibshirani (2024) provided a frequentist analysis of measures of prediction error when certain prediction models are evaluated using CV methods. These recent frequentist studies are constructive, but they focus on evaluating predictive accuracy per se. They do not study the performance of predictions in decision making, which is our concern.

### 3.1. The Fundamental Importance of the Cross-Covariate Structure of the State Space

A focus of controversy has been the asserted capacity of certain prediction methods advocated by ML researchers to successfully predict outcomes when covariate vectors have high dimension relative to sample size. Reacting to the longstanding concern that prediction becomes increasingly difficult as the dimension of the covariate vector increases (aka the curse of dimensionality), Breiman (2001a) expressed considerable optimism when he wrote (p. 209): "For decades, the first step in prediction methodology was to avoid the curse. . . . . Recent work has shown that dimensionality can be a blessing." He continued (p. 209):

> "Reducing dimensionality reduces the amount of information available for prediction. The more predictor variables, the more information. There is also information in various combinations of the predictor variables. Let's try going in the opposite direction: Instead of reducing dimensionality, increase it by adding many functions of the predictor variables."

Logically, Breiman's optimism cannot be completely realistic. The foundation of the curse of dimensionality is that, as the support of the covariate distribution grows, the probability P(x) of drawing any specified value of x into a random training sample of fixed size N necessarily decreases. Hence, the expected number of observations of y associated with any specified value of x necessarily falls. Indeed, it is zero when the covariate distribution is continuous rather than discrete. It follows that, as the support of the covariate distribution grows, prediction of y conditional on the specified value of x with a training sample of size N must increasingly rely on outcome data associated with other covariate values. However, observation of outcomes associated with other covariate values per se conveys no information about P(y|x) at the specified x-value of interest. These data become informative *only* given cross-covariate restrictions on the state space that relate P(y|x) to the conditional outcome distributions P(y|x′), x′ ≠ x.

Frequentist statistical theory has long sought to characterize how the statistical imprecision of conditional prediction varies with the maintained cross-covariate restrictions. Analysis is straightforward in parametric modeling of conditional distributions, where the distributions {P(y|x), x ∈ X}, or at least the conditional expectations {E(y|x), x ∈ X} are assumed to all be functions of a finite-dimensional parameter vector. Then cross-covariate restrictions are formalized in the specification of the parameter space. Analysis is relatively straightforward in the semiparametric approach to conditional classification called *maximum score* estimation in econometrics (Manski, 1975, 1985; Manski and Thompson, 1989) and *empirical risk minimization* in the statistical learning theory of Vapnik (1999, 2000). In both the parametric and semiparametric settings, researchers have mainly studied asymptotic questions of consistency and rates of convergence.

Analysis is more subtle, but still intuitive, in classical research on nonparametric regression, where {P(y|x), x ∈ X} or {E(y|x), x ∈ X} are assumed to vary in a smooth manner across suitably nearby values of x. The more recent body of research assuming some concept of *dimensional sparsity* is yet more subtle in that it does not a priori fix the covariate space over which the conditional distribution or mean may vary. Instead, it constrains the number of elements of the covariate vector across which variation may occur.

The ML literature using deep neural networks as predictor functions replaces dimensional sparsity with the concept of compositional sparsity. Poggio et al. (2017) wrote this (p. 503):

> "The main message is that deep networks have the theoretical guarantee, which shallow networks do not have, that they can avoid the curse of dimensionality for an important class of problems, corresponding to compositional functions, i.e., functions of functions. An especially interesting subset of such compositional functions are hierarchically local compositional functions where all the constituent functions are local in the sense of bounded small dimensionality. The deep networks that can approximate them without the curse of dimensionality are of the deep convolutional type."

Poggio and Fraser (2024) defined compositional sparsity as follows (p. 1): "The property that a compositional function have *[sic]* "few" constituent functions, each depending on only a small subset of inputs."

Thus, if the function governing variation in the outcome y with covariates x is compositionally sparse, then the claim is that a deep neural network will approximate it well. It should not be surprising that a predictor function whose structure is similar to the unknown, true function of interest will approximate it well. Recognizing this led Shamir (2020), commenting on Schmidt-Hieber (2020), to assert (p. 1912): "Essentially, we have replaced a 'curse of dimensionality' effect with a 'curse of sparsity'."

A logical follow-up question concerns whether it is realistic in practice to assume that unknown functions governing variation in y with x are compositionally sparse. Poggio et al. (2017) asked this question and conjectured a partial answer, writing (p. 517):

> "This line of arguments defines a class of algorithms that is universal and can be optimal for a large set of problems. It does not however explain why problems encountered in practice should match this class of algorithms. Though we and others have argued that the explanation may be in either the physics or the neuroscience of the brain, these arguments…are not (yet) rigorous. Our conjecture is that compositionality is imposed by the wiring of our cortex and is reflected in language. Thus, compositionality of many computations on images may reflect the way we describe and think about them."

As we see it, compositional sparsity is an intriguing concept that warrants further study. In principle, SDT can specify that a state space is composed of compositionally sparse functions and evaluate the predictive performance of deep neural networks that have this structure. However, we expect that this implementation of SDT will be computationally challenging. We also think it important to question the realism of assuming that unknown conditional means and other features of conditional distributions actually are compositionally sparse. Poggio et al. (2017) and Poggio and Fraser (2024) express optimism about this. We are less sure.

3.2. Bounded-Variation Restrictions on the State Space

In some applications, it is realistic to globally bound the variation across covariate values of a conditional mean E(y|x) or other feature of P(y|x). Focusing on the problem of identification rather than decision making with sample data, Manski and Pepper (2000) initiated study of identification with *monotone instrumental variable* assumptions, which assume that a sub-vector of x is ordered, and that E(y|x) varies monotonically across these covariate values. More general *bounded-variation* assumptions have been used to study conditional prediction of risk of illness by Manski (2018) and Li, Litvin, and Manski (2023). They have been used to study prediction of treatment response in criminal justice settings by Manski and Pepper (2013, 2018).

In settings where the outcome y is binary, bounded-variation assumptions have the simple form

$$(9) \qquad P_s(y = 1|x') + \lambda_-(x, x') \leq P_s(y = 1|x) \leq P_s(y = 1|x') + \lambda_+(x, x'), \quad \text{all } s \in S.$$

Here x is a covariate value of prediction interest, x′ is a different value, and $\lambda_-(x, x') \leq \lambda_+(x, x')$ are specified real numbers. Bounded-variation assumptions do not assert that $P_s(y = 1|x)$ varies locally smoothly with x, as in classical nonparametric regression analysis. Nor do they assume the types of cross-covariate invariance embedded in dimensional or compositional sparsity assumptions. They impose a distinct type of cross-covariate restriction on the state space.

As far as we are aware, the only research studying SDT with bounded-variation assumptions are Stoye (2012), Manski (2023), and Montiel Olea, Qiu, and Stoye (2024). These focused on the maximum regret of SDFs in certain problems of treatment choice. Stoye (2012) obtained an analytical expression for the MMR decision function when $\lambda_-(x, x') = -\lambda_+(x, x') = \kappa$, where κ is a specified positive real number that does not vary with (x, x′).

Manski (2023) considered numerical computation of maximum regret for general assumptions of form (9), when loss is an extension of the MCR where the magnitude of the loss incurred with a prediction error varies with x and s. For concreteness, the analysis considered a medical setting in which a clinician's problem is to predict illness and to use the prediction to choose between surveillance and aggressive treatment of a disease. The assumptions made about patient outcomes with each treatment implied that aggressive treatment is optimal if $P^*(y = 1|x)$ exceeds a known threshold and surveillance is optimal otherwise. This will be discussed further in Section 4.

It was supposed that, not knowing $P^*(y = 1|x)$, the clinician uses a kernel method to predict it and then acts as if the prediction is correct. A kernel estimate is a weighted average of the outcomes across different values of the covariate. Computation of maximum regret using alternative weights enables one to minimize maximum regret within the class of kernel predictors.

## 4. Probabilistic Prediction for Clinical Decision Making

Our discussion of statistical decision theory has been abstract, with mention of potential but not actual applications. We are concerned that the scarcity of applied use of SDT to evaluate predictive algorithms has been accompanied by a rapid growth in use of heuristic OOS validation methods that lack theoretical foundation. Considering the many domains in which heuristic OOS validation is becoming common, we are especially troubled by its increasing acceptance in medical research aiming to predict patient health outcomes. Probabilistic prediction of illness and treatment response is used widely to inform clinical decision making. It matters considerably to individual patients and to society at large that researchers use a well-grounded approach to assess the reliability of predictions.

Section 4.1 summarizes the trajectory of prediction methodology in medicine. Section 4.2 focuses on an important medical setting, where a clinician wants to predict illness and to use the prediction to choose between surveillance and aggressive treatment of a disease. Section 4.3 discusses a recent application of heuristic OOS validation to study this prediction and decision problem.

### 4.1. A Brief History of Probabilistic Prediction in Medicine

Empirical research on medical risk assessment has long used sample data on illness in study populations to estimate conditional probabilities of illness and to make probabilistic predictions of treatment response. Throughout the 20th century, risk assessment conditional on observed patient covariates was performed by biostatisticians who use frequentist statistical theory to propose methodology and assess findings. Certain parametric and semi-parametric models became standard. The logit model introduced by Berkson (1944) has been used to predict binary outcomes. The proportional hazards model introduced by Cox (1972) has been used to predict the timing of death and other future events. Linear regression models introduced by Galton (1886) have been used to predict the means of real-valued outcomes. Poisson regression has been used to predict outcomes taking count values (Nelder, 1974). The parameters of these models have been estimated by maximum likelihood or least squares, which have well-understood frequentist statistical properties.

Although the motivation ostensibly is to improve patient care, frequentist biostatistical analysis has commonly viewed prediction as a self-contained inferential problem rather than as a task undertaken specifically to inform treatment choice. Confidence intervals and standard errors have been used to measure statistical imprecision. When treatment choice is studied, Neyman-Pearson hypothesis tests have been used to judge whether treatment B is better than treatment A, the null hypothesis being that B is no better than A and the alternative being that it is better. These inferential methods are remote from SDT. Confidence intervals and standard errors have no obvious interpretation in the Wald theory. As discussed in Appendix A, Wald (1939) initially proposed SDT as a superior replacement for Neyman-Pearson testing.

In the 21st century, medical risk assessment is increasingly performed by computer scientists, who view prediction methods as computational algorithms rather than through the lens of statistical theory. Whereas biostatisticians have favored the use of estimation of relatively simple parametric and semi-parametric probabilistic prediction models, computer scientists fit data with increasingly complex nonparametric algorithms whose frequentist statistical properties are poorly understood.

Early on, computer scientists applied what are now considered classical nonparametric methods, such as kernel and nearest-neighbor regression; see, for example, the *Journal of Machine Learning Research* special issue on kernel methods (Cristianini et al., 2001). These are well understood from the perspective of asymptotic statistical theory. They later focused on random forest methods (Ho, 1995; Breiman, 2001b), for which some statistical theory exists. However, they have increasingly used methods whose statistical properties are hardly understood at all. We have already noted the weak state of knowledge regarding deep neural networks. Another prominent approach lacking theoretical foundation, which we have not mentioned thus far, is the Synthetic Minority Over-Sampling Technique (SMOTE) introduced by Chawla et al. (2002). Whereas frequentist biostatisticians study how methods perform across repetitions of a sampling process, computer scientists engaged in medical prediction perform K-fold or CTF OOS validation, as championed by Breiman (2002). See Deo (2015) and Rajkomar et al. (2019) for two among many review articles written for audiences of clinicians.

## 4.2. Prediction for Choice Between Surveillance and Aggressive Treatment

Choice between surveillance and aggressive treatment is a frequent clinical decision. The broad problem concerns a clinician caring for patients with observed covariates x. There are two care options for a specified disease, A denoting surveillance and B denoting aggressive treatment. The clinician must choose without knowing a patient's illness status; y = 1 if a patient is ill and y = 0 if not. Observing x, the clinician can attempt to learn the conditional probability of illness, $p_x \equiv P^*(y = 1|x)$. Medical research often proposes as-if optimization, using sample data to estimate $p_x$ and acting as if the estimate is correct.

We summarize here the simple version of this decision problem studied in Manski (2023). As there, we assume that patient welfare with care option $c \in \{A, B\}$ has the known form $U_x(c, y)$; thus, welfare varies with whether the disease occurs and with the patient covariates x. We assume that aggressive treatment is better if the disease occurs, and surveillance is better otherwise. That is,

(10a) $U_x(B, 1) > U_x(A, 1)$,

(10b) $U_x(A, 0) > U_x(B, 0)$.

The specific form of welfare function $U_x(\cdot, \cdot)$ depends on the clinical context, but inequalities (10a) – (10b) are typically realistic.

We assume that the chosen care option does not affect whether the disease occurs; hence, a patient's illness probability is simply $p_x$ rather than a function $p_x(c)$ of the care option. Treatment choice still matters because it affects the severity of illness and the patient's experience of side effects. Aggressive treatment is beneficial to the extent that it lessens the severity of illness, but harmful if it yields side effects that do not occur with surveillance.

4.2.1. Optimal Treatment Choice with Knowledge of $p_x$

Before considering decision making with sample data, suppose that the clinician knows $p_x$ and maximizes expected patient welfare conditional on x. Then an optimal decision is

(11a) Choose A if $p_x \cdot U_x(A, 1) + (1 - p_x) \cdot U_x(A, 0) \geq p_x \cdot U_x(B, 1) + (1 - p_x) \cdot U_x(B, 0)$,

(11b) Choose B if $p_x \cdot U_x(B, 1) + (1 - p_x) \cdot U_x(B, 0) \geq p_x \cdot U_x(A, 1) + (1 - p_x) \cdot U_x(A, 0)$.

The decision yields optimal expected patient welfare

(12) $\max [p_x \cdot U_x(A, 1) + (1 - p_x) \cdot U_x(A, 0),\ p_x \cdot U_x(B, 1) + (1 - p_x) \cdot U_x(B, 0)]$.

The optimal decision is easy to characterize when inequalities (10a) − (10b) hold. Let $p_x^{\#}$ denote the threshold value of $p_x$ that makes options A and B have the same expected utility. This value is

$$(13) \qquad p_x^{\#} = \frac{U_x(A, 0) - U_x(B, 0)}{[U_x(A, 0) - U_x(B, 0)] + [U_x(B, 1) - U_x(A, 1)]}$$

Option A is optimal if $p_x \leq p_x^{\#}$ and B if $p_x \geq p_x^{\#}$. Thus, optimal treatment choice does not require exact knowledge of $p_x$. It only requires knowing whether $p_x$ is larger or smaller than $p_x^{\#}$.

Manski (2023) focused on an instructive special case that occurs when aggressive treatment neutralizes disease, in the sense that $U_x(B, 0) = U_x(B, 1)$. Let $U_{xB}$ denote welfare with aggressive treatment. Then (10) – (13) reduce to

$$(14) \quad U_x(A, 0) > U_{xB} > U_x(A, 1).$$

$$(15a) \quad \text{Choose A if } p_x \cdot U_x(A, 1) + (1 - p_x) \cdot U_x(A, 0) \geq U_{xB},$$

$$(15b) \quad \text{Choose B if } U_{xB} \geq p_x \cdot U_x(A, 1) + (1 - p_x) \cdot U_x(A, 0).$$

$$(16) \quad \max\, [p_x \cdot U_x(A, 1) + (1 - p_x) \cdot U_x(A, 0), U_{xB}].$$

$$(17) \qquad p_x^{\#} = \frac{U_x(A, 0) - U_{xB}}{U_x(A, 0) - U_x(A, 1)}.$$

Further simplification occurs when one normalizes the location and scale of welfare by setting $U_x(A, 0) = 1$ and $U_x(A, 1) = 0$. Then $1 > U_{xB} > 0$ and $p_x^{\#} = 1 - U_{xB}$. We consider this special case below.

4.2.2. Maximum Regret of As-If Optimization

Now suppose that the clinician does not know $p_x$. The state space lists all feasible values of $p_x$. In the abstract notation of Section 2, the objective function $w(\cdot, \cdot): C \times S \rightarrow R^1$ has this form: $w(A, s) = p_{sx} \cdot U_x(A, 1) + (1 - p_{sx}) \cdot U_x(A, 0)$ and $w(B, s) = U_{xB}$.

Suppose that the clinician does not know whether $p_x$ is smaller or larger than $p_x^{\#} = 1 - U_{xB}$. Formally, suppose the clinician knows that $p_{mx} < 1 - U_{xB} < p_{Mx}$, where $p_{mx} \equiv \min_{s \in S} p_{sx}$ and $p_{Mx} \equiv \max_{s \in S} p_{sx}$. Then the clinician cannot maximize expected patient welfare conditional on x.

The clinician can, however, use sample data to estimate $p_x$ and then act as if the estimate is correct. Let $\Psi$ be a sample space and let $Q_s$ be a sampling distribution with realizations $\psi \in \Psi$. Let $\varphi_x(\psi)$ be a point estimate of $p_x$. The clinician can maximize expected welfare acting as if $p_x = \varphi_x(\psi)$. Then the chosen care option is A if $\varphi_x(\psi) \leq 1 - U_{xB}$ and is B if $\varphi_x(\psi) > 1 - U_{xB}$.

Now consider the regret of treatment choice acting as if $p_x = \varphi_x(\psi)$. Let $e[p_{sx}, \varphi_x(\psi), U_{xB}]$ denote the occurrence of an error in state s when $\varphi_x(\psi)$ is used to choose treatment. That is, $e[p_{sx}, \varphi_x(\psi), U_{xB}] = 1$ when $p_{sx}$ and $\varphi_x(\psi)$ yield different treatments, while $e[p_{sx}, \varphi_x(\psi), U_{xB}] = 0$ when $p_{sx}$ and $\varphi_x(\psi)$ yield the same treatment. Regret using estimate $\varphi_x(\psi)$ is

$$
\begin{aligned}
(18) \quad R_{sx}[\varphi_x(\psi)] &= \max(1 - p_{sx}, U_{xB}) - (1 - p_{sx}) \cdot 1[1 - \varphi_x(\psi) \geq U_{xB}] - U_{xB} \cdot 1[U_{xB} > 1 - \varphi_x(\psi)] \\
&= |(1 - p_{sx}) - U_{xB}| \cdot 1[1 - p_{sx} \geq U_{xB} > 1 - \varphi_x(\psi) \text{ or } 1 - p_{sx} < U_{xB} \leq 1 - \varphi_x(\psi)] \\
&= |(1 - p_{sx}) - U_{xB}| \cdot e[p_{sx}, \varphi_x(\psi), U_{xB}].
\end{aligned}
$$

Expected regret across repeated samples is

$$
(19) \quad E_s\{R_{sx}[\varphi_x(\psi)]\} = |(1 - p_{sx}) - U_{xB}| \cdot Q_s\{e[p_{sx}, \varphi_x(\psi), U_{xB}] = 1\}.
$$

Maximum expected regret across the state space is $\max_{s \in S} E_s\{R_{sx}[\varphi_x(\psi)]\}$.

---

Algorithm for calculation of maximum regret

1. Choose a grid on S
2. For each s in the grid, calculate expected regret (19)
    a. Compute $|(1 - p_{sx}) - U_{xB}|$
    b. Use Monte Carle integration to approximate $Q_s\{e[p_{sx}, \varphi_x(\psi), U_{xB}] = 1\}$ (see Section 2.1.3)
    c. Multiply a × b
3. Maximize expected regret over S

---

Manski (2023) applied this algorithm to numerical computation of maximum regret with as-if optimization when bounded-variation assumptions restrict the state space and $\varphi_x(\psi)$ is a kernel estimate of $p_x$. In the present paper, Section 4.2.3 considers computation when $\varphi_x(\psi)$ is an estimate that uses CV to choose tuning parameters.

#### 4.2.3. Illustration: Maximum Regret of As-If Optimization with Cross-Validated Estimation of $p_x$

We earlier mentioned the recent contribution of Bates, Hastie, and Tibshirani (2024), who performed a frequentist analysis of measures of prediction error when certain prediction models are evaluated using CV methods. Their study focused on the prevalent ML use of CV to evaluate predictive accuracy. They did not address the distinct use of CV to choose tuning parameters in predictive statistical models, such as the bandwidth in kernel estimation or the penalty in LASSO estimation. They wrote (p. 1436):

> "CV is often used to compare predictive models, such as when selecting a model or a good value of a learning algorithm's hyperparameters . . . . While we expect that our proposed estimator would be of use for hyperparameter selection because it yields more accurate confidence intervals for prediction error, we do not pursue this problem further in the present work."

From the perspective of decision making using sample data, we question their conjecture relating hyperparameter selection to accuracy of confidence intervals for prediction error. We noted earlier that confidence intervals have no obvious interpretation in SDT. We recommend evaluation of CV choice of tuning parameters by the maximum regret of the resulting decisions.

To illustrate the striking difference between the type of analysis performed by Bates *et al.* and what we have in mind, consider the example they present in their opening Section 1.1. They write (p. 1434):

> "we consider a sparse logistic regression model . . . . with $n$ = 90 observations of $p$ = 1000 features, and a coefficient vector $\theta = c \cdot (1, 1, 1, 1, 0, 0, \ldots)^T \in \mathrm{R}^p$ with four nonzero entries of equal strength. The feature matrix $X \in \mathrm{R}^{n \times p}$ is comprised of iid standard normal variables, and we chose the signal strength $c$ so that the Bayes misclassification rate is 20%. We estimate the parameters using $\ell_1$-penalized logistic regression with a fixed penalty level. In this case, naïve confidence intervals for prediction error are far too small: intervals with desired miscoverage of 10% give 31% miscoverage in our simulation. . . . . The intervals need to be made larger by a factor of about 1.6 to obtain coverage at the desired level in this case.

Their reference to the "Bayes misclassification rate" indicates that their concern is to measure the expected frequency of misclassification across the distribution of $X$ when the estimated logit probability is used to predict the binary event, the prediction being 1 if the estimated probability is greater than ½ and 0 otherwise. With their assumption that $X$ is iid standard normal and that the binary outcome is generated by the specified sparse logit model, they state that the oracle Bayes misclassification rate is 20%. Using sample data to estimate the logit model yields a larger misclassification rate across repeated samples, which they seek to characterize using CV.

The nature of their analysis differs in numerous ways from the computation of maximum regret that we described in Section 4.2.2. There we considered treatment choice for patients with a specified value of $X$, rather than across the distribution of $X$. In a specified state of nature, expected regret across repeated samples was given in (19), which multiplies the magnitude of a treatment error times the probability of an error. Calculation of maximum regret across the state space recognizes that, in practice, the true distribution of $X$ and of the binary outcome conditional on $X$ are not known, whereas Bates *et al.* presume them to be iid normal and given by a particular sparse logistic model. Finally, we would envision using CV to choose the penalization parameter in the $\ell_1$-penalized logistic regression whereas they fix the penalty level.

4.3. A Machine Learning Model of the Probability of a Coronary Blockage

It should by now be evident that evaluation of an estimate $\varphi_x(\cdot)$ of $p_x$ by maximum regret differs fundamentally from ML evaluation by ex-post prediction accuracy. In the CTF paradigm, one uses a training sample $\psi_{tr}$ to compute $\varphi_x(\psi_{tr})$ and then uses this estimate to predict illness in a test sample $\psi_{test}$. A standard practice is to predict y = 1 if $\varphi_x(\psi_{tr})$ exceeds a specified threshold, say $\gamma$, and y = 0 otherwise. Prediction accuracy is measured by the *positive predictive value*, the fraction of the test sample with y = 1 conditional on $\varphi_x(\psi_{tr}) \geq \gamma$, and the *negative predictive value*, the fraction of the test sample with y = 0 conditional on $\varphi_x(\psi_{tr}) \leq \gamma$. These measures of ex post prediction accuracy do not consider performance across samples or over the state space. Nor do they consider the patient welfare achieved when an estimate is used to choose a treatment.

An ambitious application of heuristic OOS evaluation of ML to choice between surveillance and aggressive treatment occurs in Mullainathan and Obermeyer (2022), who study the decision faced by an emergency room physician when a patient presents with symptoms that may indicate a heart attack. In this case, aggressive treatment begins by ordering an invasive and costly test, such as cardiac catheterization, to determine whether a new blockage in the coronary arteries exists and, if so, the location of the blockage. Mullainathan and Obermeyer specify a blockage-probability threshold model to determine whether or not to order the test for a blockage. They use data on almost 250,000 emergency room visits to estimate an ML model of the blockage probability conditional on over 16,000 patient covariates. Considering the information available to physicians who make the testing decision and the incentives they face, the authors seek to determine whether actual testing decisions deviate from those that would be optimal with their model. They use (p. 679): "actual health outcomes to evaluate whether those deviations represent mistakes or physicians' superior knowledge."

The ML method they employ is described as an "ensemble" model that combines health outcome predictions of four different models. Two algorithms—gradient boosted trees and LASSO—are each trained on two subsets of the data—tested patients and untested patients. The parameters of the four models are each tuned using 5-fold CV in the same 70-percent training sample. The predictions of the four models are then used as the four covariates in a logit regression that predicts patient outcomes in a separate 5-percent ensemble training sample The remaining 25-percent holdout sample is used to identify cases where low-probability patients were tested but the finding was negative (called "overtesting") and high-probability patients were not tested but subsequently have "major adverse cardiac events" (called "undertesting").

The methods used in this study exemplify the current acceptance and growing use of heuristic OOS validation methods that lack theoretical foundation. These validation methods differs sharply from how one would proceed using SDT to study the choice between surveillance and aggressive treatment, as discussed in Section 4.2. Application of SDT to high-dimensional covariate contexts and complex modeling approaches such as in the Mullainathan-Obermeyer study is not currently computationally tractable. We see this as a challenge to be overcome.

## 5. Conclusion

Whereas statisticians and econometricians have used frequentist or Bayesian statistical theory to propose and evaluate prediction methods, ML researchers have largely performed heuristic K-fold and CTF OOS validation. These types of OOS evaluation cannot yield generalizable lessons.

Considering our critique of OOS evaluation, a reviewer of an earlier draft of this paper commented: "while it is clear that the authors believe modern machine learning lacks a particular style of rigor to their taste, it is also true that the field has produced breakthroughs that would have been unimaginable 15 years ago." The reviewer then provided an impressive list of examples.

We agree that the capacity to perform conditional prediction of events in stable populations has improved remarkably recently, but we do not see a basis to attribute the improvement to OOS evaluation of predictive algorithms. Our perception is that the improvements stem mainly from the combination of (1) vast increases in the availability of data to train predictive models, (2) vast increases in the capacity of computer hardware to perform prediction rapidly, and (3) addition of new approaches to the body of nonparametric regression methods developed in statistics from the 1950s onward, perhaps most notably the methodology of deep neural networks for approximation of functions with compositional sparsity.

It may be that the heuristic simplicity of OOS evaluation has contributed to the adoption of ML predictive algorithms, but we don't see how it has directly contributed to the improvements themselves. To the contrary, we are concerned that selective publication of purportedly successful OOS evaluations may yield a misleadingly positive impression of some algorithms and encourage their premature adoption. We see this as particularly problematic in medical risk assessment.

In Sections 1 through 3, we argued abstractly that ML researchers should instead perform comprehensive OOS evaluation using SDT. SDT is remote from the heuristic types of OOS evaluation currently practiced, but it is not remote conceptually. It performs OOS evaluation across all possible (1) training samples, (2) populations that may generate training data, and (3) populations of prediction interest.

SDT is simple in abstraction, but it is often computationally demanding to implement. Some progress has been made in tractable implementation of SDT in prediction problem with low-dimensional covariates. We recognize that extending the approach to prediction problems with high-dimensional covariates may be computationally challenging. Yet we should confront the challenge. We are very concerned about potentially harmful consequences, especially in clinical decision making but also in many other settings, arising from the use of predictive algorithms evaluated by K-fold or CTF validation rather than by comprehensive OOS evaluation using SDT. We therefore call on ML researchers to join with econometricians and statisticians in expanding the domain within which implementation of SDT is tractable.

Appendix A: Binary Classification Problems

Binary classification problems are ones in which the choice set C contains two feasible actions, say A and B. An SDF c(·) partitions Ψ into two regions that separate the data yielding choice of each action. These are $\Psi_{c(\cdot)A} \equiv [\psi \in \Psi\text{: } c(\psi) = A]$ and $\Psi_{c(\cdot)B} \equiv [\psi \in \Psi\text{: } c(\psi) = B]$. Thus, one chooses A if the data lie in set $\Psi_{c(\cdot)A}$ and B if the data lie in $\Psi_{c(\cdot)B}$.

Choice between two actions can be viewed as hypothesis tests, but the Wald perspective on testing differs from that of Neyman and Pearson (1928, 1933). Wald (1939) proposed evaluation of the performance of an SDF for binary classification by the expected loss that it yields across realizations of the sampling process. The loss distribution in state s is Bernoulli, with mass points max [L(a, s), L(b, s)] and min [L(a, s), L(b, s)]. These coincide if L(a, s) = L(b, s). When L(a, s) ≠ L(b, s), let $R_{c(\cdot)s}$ denote the probability that c(·) yields an error, choosing the inferior action over the superior one. That is,

$$\text{(A1)} \qquad R_{c(\cdot)s} = Q_s[c(\psi) = b] \quad \text{if } L(a, s) < L(b, s),$$
$$= Q_s[c(\psi) = a] \quad \text{if } L(b, s) < L(a, s).$$

These are the probabilities of Type I and Type II errors.

The probabilities that loss equals min [L(a, s), L(b, s)] and max [L(a, s), L(b, s)] are $1 - R_{c(\cdot)s}$ and $R_{c(\cdot)s}$. Hence, expected loss is

$$\text{(A2)} \qquad E_s\{L[c(\psi), s]\} = R_{c(\cdot)s}\{\max [L(a, s), L(b, s)]\} + [1 - R_{c(\cdot)s}]\{\min [L(a, s), L(b, s)]\}$$
$$= \min [L(a, s), L(b, s)] + R_{c(\cdot)s}\cdot|L(a, s) - L(b, s)|.$$

$R_{c(\cdot)s}\cdot|L(a, s) - L(b, s)|$ is the expected regret of c(·). Thus, expected regret, defined in abstraction in (6), has a simple form when choice is binary. It is the product of the error probability and the magnitude of the incremental loss when an error occurs.

It is particularly simple to determine the regret of a point predictor $p(\cdot)$ of a binary outcome y, when measuring loss by mean square error (MSE) or the misclassification rate (MCR). Let y be generated by a Bernoulli distribution P(y), $\psi$ be generated by a sampling distribution $Q(\psi)$, and let the realizations of y and $\psi$ be statistically independent.

When loss is MSE and the point predictor $p(\cdot)$ can take any real value, the risk of $p(\cdot)$ in state s is

$$\text{(A3)} \qquad E[y - p(\psi)]^2 = V_s(y) + V_s[p(\psi)] + \{E_s(y) - E_s[p(\psi)]\}^2.$$

Risk in state s is minimized by setting $p(\psi) = E_s(y)$ for all data realizations $\psi$, yielding the minimum risk value $V_s(y)$. Hence, the regret of $p(\cdot)$ in state s is $V_s[p(\psi)] + \{E_s(y) - E_s[p(\psi)]\}^2$.

When y is binary and the point predictor $p(\cdot)$ is constrained to be binary, $V_s[p(\psi)] = Q_s[p(\psi) = 1] - Q_s[p(\psi) = 1]^2$, $E_s(y) = P_s(y = 1)$, and $E_s[p(\psi)] = Q_s[p(\psi) = 1]$. Hence, regret is

$$\begin{aligned}
\text{(A4)} \quad & Q_s[p(\psi) = 1] - Q_s[p(\psi) = 1]^2 + \{P_s(y = 1) - Q_s[p(\psi) = 1]\}^2 \\
&= Q_s[p(\psi) = 1] - Q_s[p(\psi) = 1]^2 + P_s(y = 1)^2 + Q_s[p(\psi) = 1]^2 - 2\cdot P_s(y = 1)\cdot Q_s[p(\psi) = 1] \\
&= Q_s[p(\psi) = 1][1 - P_s(y = 1)] + P_s(y = 1)\{P_s(y = 1) - Q_s[p(\psi) = 1]\} \\
&= Q_s[p(\psi) = 1][1 - P_s(y = 1)] + P_s(y = 1)\{1 - Q_s[p(\psi) = 1]\} - P_s(y = 1)[1 - P_s(y = 1)].
\end{aligned}$$

The first term is the probability of a misclassification of the form $[y = 0, p(\psi) = 1]$. The second term is the probability of a misclassification of the form $[y = 1, p(\psi) = 0]$. The third term is the outcome variance $V_s(y)$.

When loss is MCR, the risk of $p(\cdot)$ in state s is

$$\text{(A5)} \qquad P_sQ_s[y \neq p(\psi)] = Q_s[p(\psi) = 1][1 - P_s(y = 1)] + P_s(y = 1)\{1 - Q_s[p(\psi) = 1]\}.$$

Risk in state is minimized by setting $p(\psi) = 1$ for all $\psi$ if $P_s(y = 1)] > ½$ and $p(\psi) = 0$ for all $\psi$ if $P_s(y = 1)] \leq ½$, yielding the minimum risk value $\min[P_s(y = 1), 1 - P_s(y = 1)]$. Hence, regret is

(A6) $$Q_s[p(\psi) = 1][1 - P_s(y = 1)] + P_s(y = 1)\{1 - Q_s[p(\psi) = 1]\} - \min[P_s(y = 1), 1 - P_s(y = 1)].$$

Thus, regret when loss is MSE or MCR differs only in that the former subtracts $P_s(y = 1)[1 - P_s(y = 1)]$ from the probability of misclassification and the latter subtracts $\min[P_s(y = 1), 1 - P_s(y = 1)]$. The two differ because the MSE calculation permits the prediction $p(\cdot)$ to take any real value, whereas the MCR calculation constrains the prediction to take the value 0 or 1.